\documentclass[prl,twocolumn,showpacs,floatfix]{revtex4}
\usepackage[dvips]{graphicx}
\usepackage{hyperref}
\usepackage{bm}
\usepackage{dsfont}
\usepackage{rotating}
\usepackage{times}
\usepackage{color}

\begin{document}
\def\diff{\mathrm{d}}

\def\jpb{J. Phys. B: At. Mol. Opt. Phys.~}
\def\pra{Phys. Rev. A~}
\def\prb{Phys. Rev. B~}
\def\prl{Phys. Rev. Lett.~}
\def\jmo{J. Mod. Opt.~}
\def\jetp{Sov. Phys. JETP~}
\def\etal{{\em et al.}}

\def\pabl#1#2{\frac{\partial #1}{\partial #2}}
\def\abl#1#2{\frac{\diff #1}{\diff #2}}

\def\varphit{\tilde{\varphi}}
\def\Mtilde{\tilde{M}}
\def\Ktilde{\tilde{K}}
\def\gammatilde{\tilde{\gamma}}

\def\twovec#1#2{\left( \begin{array}{c} #1 \\ #2 \end{array}\right)}
\def\twotimestwo#1#2#3#4{\left( \begin{array}{cc} #1 & #2 \\ #3 & #4 \end{array}\right)}

\def\vekt#1{{\bm{#1}}}
\def\vect#1{\vekt{#1}}
\def\vektalpha{\vekt{\alpha}}
\def\vekta{\vekt{a}}
\def\vektr{\vekt{r}}
\def\vektd{\vekt{d}}
\def\vektx{\vekt{x}}
\def\vektp{\vekt{p}}
\def\NRNOmax{N_{\mathrm{RNO}}}
\def\Ncyc{N_{\mathrm{cyc}}}
\def\Vn{V_{\mathrm{n}}}
\def\Vi{V_{\mathrm{i}}}
\def\Hi{H_{\mathrm{i}}}
\def\vee{v_{\mathrm{ee}}}
\def\ee{{\mathrm{ee}}}
\def\fee{f_{\mathrm{ee}}}
\def\vektpc{\vekt{p}_{\mathrm{c}}}
\def\vektE{\vekt{E}}
\def\vekte{\vekt{e}}
\def\vektA{\vekt{A}}
\def\vektEhat{\hat{\vekt{E}}}
\def\vektB{\vekt{B}}
\def\vektv{\vekt{v}}
\def\vektk{\vekt{k}}
\def\vektkhat{\hat{\vekt{k}}}
\def\reff#1{(\ref{#1})}

\def\calH{{\cal H}}
\def\calP{{\cal P}}
\def\calD{{\cal D}}
\def\calh{{\cal h}}
\def\calM{{\cal M}}
\def\calK{{\cal K}}

\def\Tr{{\mathrm{Tr}}}
\def\Up{U_{\mathrm{p}}}
\def\Ip{I_{\mathrm{p}}}
\def\Tp{T_{\mathrm{p}}}
\def\Ss{S_{\vektp\mathrm{s}}}
\def\SIs{S_{\vektp\Ip\mathrm{s}}}
\def\Cps{C_{\vektp\mathrm{s}}}
\def\Cpnulls{C_{\vektp_0\mathrm{s}}}
\def\SI{S_{\vektp\Ip}}
\def\SIsnull{S_{\vektp_0\Ip\mathrm{s}}}
\def\zt{z_{\mathrm{t}}}
\def\ts{t_{\vektp\mathrm{s}}}
\def\tsnull{t_{\vektp_0\mathrm{s}}}
\def\tnulltilde{\tilde{t}_0}
\def\omegap{\omega_p}
\def\omegaMie{\omega_M}
\def\imagi{\mathrm{i}}
\def\eulere{\mathrm{e}}

\def\halb{\frac{1}{2}}

\def\phitilde{\tilde{\phi}}
\def\alphatilde{\tilde{\alpha}}
\def\phitildeperp{\tilde{\phi}^{\perp}}

 \def\psihat{{\hat{\psi}}}
 \def\psihatdag{{\hat{\psi}^\dagger}}
\def\beq{\begin{equation}}
\def\eeq{\end{equation}}

\def\energy{{\cal E}}

\def\Ehat{\hat{E}}
\def\Ahat{\hat{A}}
\def\ahat{\hat{a}}
\def\ahatdag{\hat{a}^\dagger}

\def\ket#1{\vert #1\rangle}
\def\bra#1{\langle#1\vert}
\def\braket#1#2{\langle #1 \vert #2 \rangle}

\def\makered#1{{\color{red} #1}}
\def\makegreen#1{{\color{green} #1}}

\def\Re{\,\mathrm{Re}\,}
\def\Im{\,\mathrm{Im}\,}

\def\varphic{\varphi_{\mathrm{c}}}

\def\tini{t_\mathrm{i}}
\def\tfinal{t_\mathrm{f}}
\def\vxc{v_\mathrm{xc}}
\def\vextop{\hat{v}_\mathrm{ext}}
\def\VC{V_\mathrm{c}}
\def\VHX{V_\mathrm{Hx}}
\def\VHXC{V_\mathrm{Hxc}}
\def\wop{\hat{w}}
\def\Gammaevenodd{\Gamma^\mathrm{even,odd}}
\def\Gammaeven{\Gamma^\mathrm{even}}
\def\Gammaodd{\Gamma^\mathrm{odd}}

\def\Hop{\hat{H}}
\def\hop{\hat{h}}
\def\Cop{\hat{C}}
\def\HopKS{\hat{H}_\mathrm{KS}}
\def\HKS{H_\mathrm{KS}}
\def\Top{\hat{T}}
\def\TopKS{\hat{T}_\mathrm{KS}}
\def\VopKS{\hat{V}_\mathrm{KS}}
\def\VKS{{V}_\mathrm{KS}}
\def\vKS{{v}_\mathrm{KS}}
\def\Ttildeop{\hat{\tilde{T}}}
\def\Ttilde{{\tilde{T}}}
\def\Vextop{\hat{V}_{\mathrm{ext}}}
\def\Vext{V_{\mathrm{ext}}}
\def\Vopee{\hat{V}_{{ee}}}
\def\psiopdag{\hat{\psi}^{\dagger}}
\def\psiop{\hat{\psi}}
\def\vext{v_{\mathrm{ext}}}
\def\Vee{V_{ee}}
\def\nop{\hat{n}}
\def\Uop{\hat{U}}
\def\Wop{\hat{W}}
\def\bop{\hat{b}}
\def\bopdag{\hat{b}^{\dagger}}
\def\qop{\hat{q}}
\def\jop{\hat{j\,}}
\def\vHxc{v_{\mathrm{Hxc}}}
\def\vHx{v_{\mathrm{Hx}}}
\def\vH{v_{\mathrm{H}}}
\def\vc{v_{\mathrm{c}}}
\def\xop{\hat{x}}

\def\Wcmcm{W/cm$^2$}

\def\varphiexact{\varphi_{\mathrm{exact}}}

\def\fmathbox#1{\fbox{$\displaystyle #1$}}

\title{Two-Color Strong-Field Photoelectron Spectroscopy and the Phase of the Phase}

\author{S.~Skruszewicz}
\affiliation{Institut f\"ur Physik, Universit\"at Rostock, 18051 Rostock, Germany}

\author{J.~Tiggesb\"aumker}
\affiliation{Institut f\"ur Physik, Universit\"at Rostock, 18051 Rostock, Germany}

\author{K.-H.~Meiwes-Broer}
\affiliation{Institut f\"ur Physik, Universit\"at Rostock, 18051 Rostock, Germany}

\author{M.~Arbeiter}
\affiliation{Institut f\"ur Physik, Universit\"at Rostock, 18051 Rostock, Germany}

\author{Th.~Fennel}
\affiliation{Institut f\"ur Physik, Universit\"at Rostock, 18051 Rostock, Germany}

\author{D.~Bauer}
\affiliation{Institut f\"ur Physik, Universit\"at Rostock, 18051 Rostock, Germany}

\date{\today}

\begin{abstract} 
The presence of a weak second-harmonic field in an intense-laser ionization experiment affects the momentum-resolved electron yield, depending on the relative phase  between the $\omega$ and the $2\omega$ component. The proposed two-color 'phase-of-the-phase spectroscopy' quantifies for each final electron momentum a relative-phase contrast (RPC) and a 'phase of the phase' (PP), describing how much and with which phase lag, respectively, the yield changes as function of relative phase.  Experimental results for RPC and PP spectra for rare gas atoms and CO$_2$ are presented. The spectra demonstrate a rather universal structure that is analyzed with the help of a simple model based on electron trajectories, wave-packet spreading, and (multiple) rescattering.  
Details in the PP and RPC spectra are target sensitive and thus may be used to extract structural (or even dynamical) information with high accuracy.

\end{abstract}
\maketitle

Momentum-resolved photoelectron spectra from strong-field ionization of atoms and molecules contain a wealth of information about the ionizing laser field, the target, and ultrafast processes that may occur in it (see, e.g., Refs.~\cite{schultzvrakking, puils} for recent overviews).
 If one is able to disentangle this information one may use  photoelectron spectra to image the entire ionization dynamics and the structural information convoluted into it. Thanks to the appealing possibility to analyze most of strong-field ionization dynamics in terms of semi-classical electron trajectories it  has been demonstrated that many of the complex and, at first sight, puzzling features could finally be explained in simple and intuitive terms. Examples are various low-energy structures \cite{ref_Nature_Phys_LES,ref_PRL_Quan,liu,ref_PRL_TCSFA_LES,kaestner,lemell,dura,moellerbecker}, intra and inter-cycle interferences \cite{ref_arbo_inter_intra,bian,xmtong}, ``holographic side-lobes'' and the role of multiple returns \cite{ref_Science_Huismans,ref_PRL_huismansII,hickstein}, molecular strong-field ionization \cite{meckel,odenweller}, or ``interference carpets'' \cite{ref_PRL_perpendicular_carpet}.

Refined strong-field experiments make use of additional experimental ``knobs,''  such as a pump-probe delay, a carrier-envelope phase (CEP) in the case of few-cycle pulses  (see, e.g., Refs.~\cite{paulus} and  \cite{krausz,ref_JPB_review} for reviews), or a relative phase between two laser fields of different frequency  \cite{mauritsson09,dahlstr2011,ray,shafir,xie,arbo,doerner2015}. The resulting changes in the photoelectron or, complementary, high-harmonics spectra  may then help to 
unequivocally indentify certain ionization scenarios. In this Letter, we present experimental results obtained with a co-linear, two-color laser set-up.

Consider the measurement of an observable  (here the electron yield). If this observable depends on a tunable parameter (here the relative phase between the $\omega$ and the $2\omega$ pulse), a measurement of the change of the observable as function of the parameter reveals additional information about the underlying physical mechanism (here the target and laser-sensitive ionization dynamics). A general question is how to represent this additional information. If the parameter is periodic (like the relative phase) a Fourier-transform of the observable seems adequate. In our case, the momentum-resolved photoemission signal is Fourier-transformed with respect to the relative phase. The absolute value of the complex Fourier transform gives a  relative-phase-contrast (RPC), its phase the 'phase of the phase' (PP). In this work, we present experimental RPC and PP spectra for various targets and analyze their common features in terms of ``simple man's theory'' (SMT). The main findings are: (i) the overall structure of the PP spectra is largely target-independent and displays features that can be assigned to certain electron trajectories; (ii) target-dependent features are clearly visible in the RPC and PP spectra, thus making two-color PP spectroscopy an attractive approach for revealing structural information.

The basic structure of ``ordinary'' photoelectron spectra is qualitatively well understood. For linear laser polarization, the so-called ``direct electrons'' extend up to energies $p_z^2/2=A_0^2/2 = 2 \Up$, where $p_z$ is the photoelectron momentum along the polarization direction, $A_0$ is the vector potential amplitude, and $\Up=A_0^2/4$ is the ponderomotive energy (atomic units are used unless otherwise stated). If electrons are driven back to their parent ion by the laser field, energies up to  $10 \Up$ may occur upon rescattering. 
A wealth of information is encoded in this high-energy above-threshold ionization (HATI) part of the photoelectron spectra about both (i) the driving laser field and (ii) the structure of the target \cite{spanner04,blaga2012} (via its scattering cross section \cite{morishita,chen}). Moreover, HATI is more robustly accessible to theoretical modeling than the low-energy part of photoelectron spectra, which is plagued by the necessity to take Coulomb-corrections into account \cite{ref_JMO_Popruzhenko,ref_Coulomb_asymmetry_circular_CCSFA,ref_EVA,torlinaI,ref_DDCV_Arbo_Ciappina,tmy_latest}.

Bichromatic
$\omega$-$2\omega$-pulses with parallel polarizations of the field components and adjustable relative phase $\varphi$ were generated with a set-up similar to the one in Ref.~\cite{DudNPhys06}. 
Briefly, a
Ti:sapphire laser system 
provides
$100$-fs pulses at $794$\,nm. The second harmonic is generated in a
100-$\mu$m thick 
BBO~I
crystal. 
The
$\omega$-$2\omega$ intensity ratio is controlled by detuning the phase
matching conditions through a tilt of the crystal. 
Birefringent calcite crystals compensate for the time lag between the $\omega$ and
 $2\omega$ pulse.
 The relative phase lag is controlled by
two glass wedges mounted on  piezo-driven motors. 
 The laser pulses are focused
into the extraction region of a homebuilt high-energy
velocity-map-imaging (VMI) spectrometer~\cite{SkrIJMS14} by a concave
silver mirror having a focal length of $300$~mm. 
Photoemission from Xe was used to optimize the pulse overlap.
 The MCP back plate of the detector system is
gated using a fast electronic switch
to suppress spurious signals.

Our two-color field is described by the vector potential 
\beq  \vektA(t) = A_0 \vekte_z [\sin\omega t + \xi \sin(2\omega t + \varphi) ] \label{eq:vecpot} \eeq
in dipole approximation.
The use of long pulses ensures that envelope effects are  unimportant. Throughout this work the $2\omega$ field is  kept weak ($\xi\ll 1$).

\begin{figure}
\includegraphics[width=0.8\columnwidth]{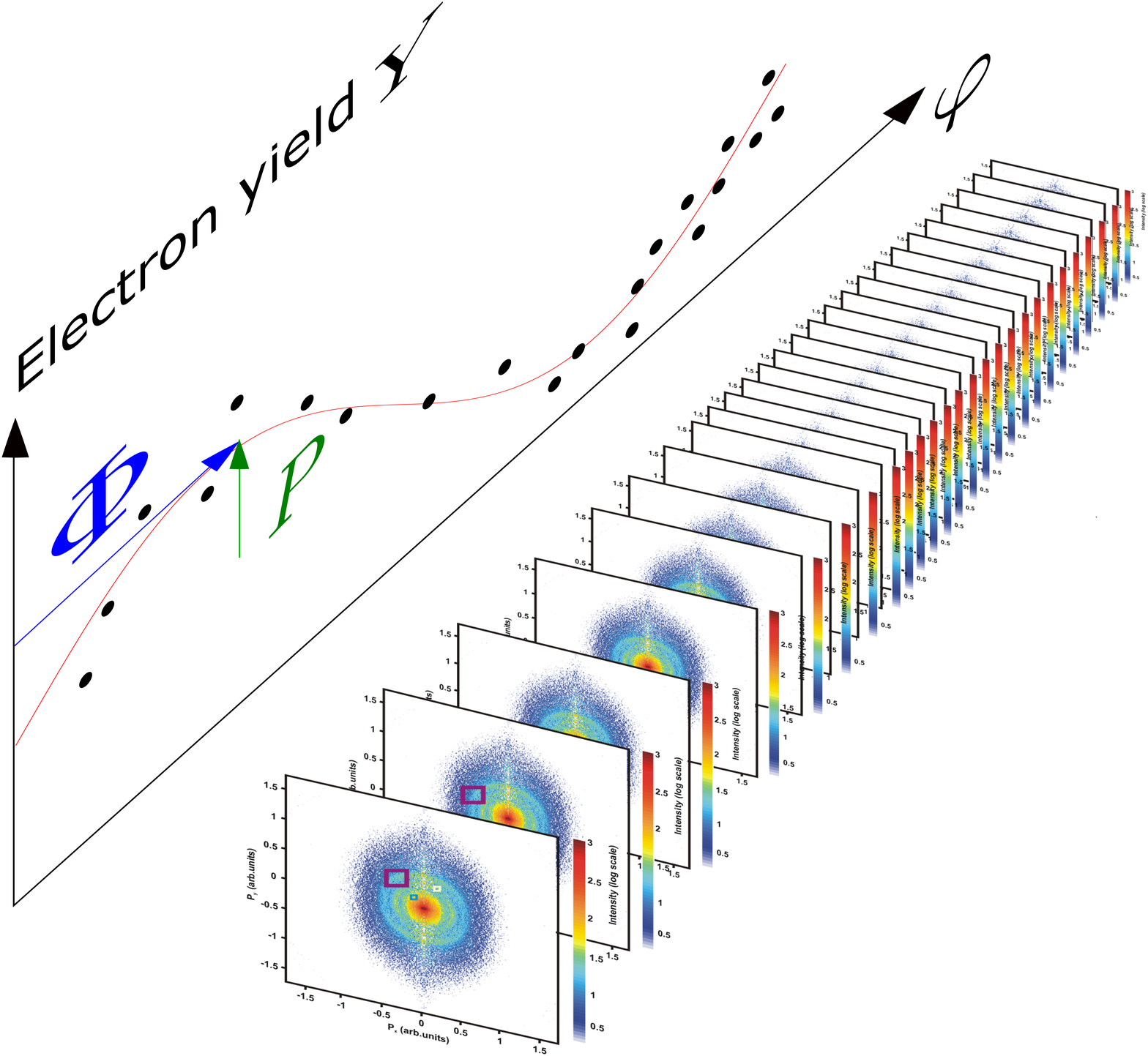} 
\caption{(color online). Schematic illustration of RPC $P$ and PP $\Phi$. Given a sequence of photoelectron momentum distributions (as obtained with the VMI spectrometer) for varied relative phase $\varphi$, the change in electron yield  (dots) for given photoelectron momentum (small square in the spectra) is fitted by $P \cos(\varphi+\Phi)$ (red curve). 
 \label{fig:VMI} }
\end{figure}

Figure~\ref{fig:VMI} introduces schematically the two quantities---RPC $P$ and PP $\Phi$---used to describe how the photoelectron yield $Y(\varphi)$ changes as function of relative phase. For each final momentum $p_z$, $p_x$, the fundamental change in the yield, i.e.,  $Y_1(\varphi)=P\cos(\varphi+\Phi)$, is extracted via the Fourier analysis mentioned above. The fundamental ($N=1$) component is found to dominate over higher-harmonic contributions $Y_N$, $N>1$ for most final momenta. A similar analysis has been performed previously with respect to the CEP for  SiO$_2$ nanospheres \cite{zhereb}.

\begin{figure}
\includegraphics[width=0.9\columnwidth]{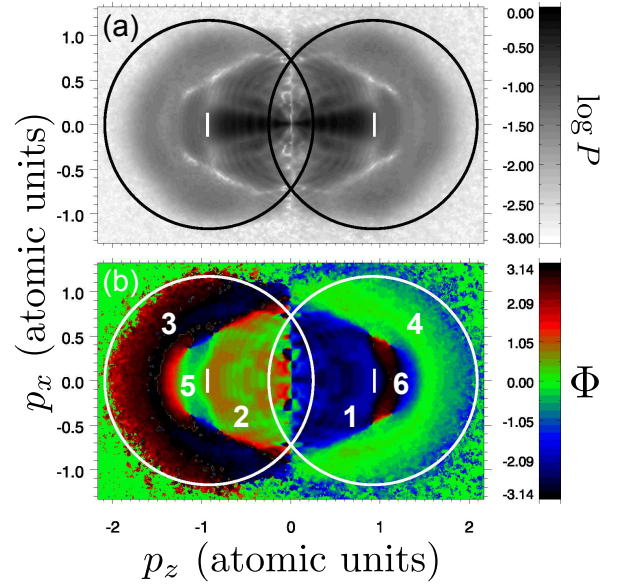} 
\caption{(color online).  Experimental RPC (a) and PP (b) spectra for Ar, calculated from $66$ Abel-projected VMI spectra per $\varphi$ interval $[0,2\pi]$. The 794~nm / 397-nm components of the $100$-fs two-color pulse had intensities $\simeq 10^{14}/10^{12}$\,\Wcmcm, respectively (i.e., $\xi=0.05$). In (a), $P$ has been normalized to $\max P$.  Circles and vertical lines indicate $10\Up$ rescattering rings and $2\Up$ cut-offs, respectively. White numbers in (b) are referred to in text.  
 \label{fig:Ar} }
\end{figure}

Experimentally determined RPC and PP spectra for Ar are presented in Fig.~\ref{fig:Ar}. The RPC in 
Fig.~\ref{fig:Ar}a shows that the direct electrons vary most with $\varphi$ (black and dark gray), while the rescattered electrons vary with about one order of magnitude less contrast.  The direct electrons with $p_z>0$ in the PP spectrum in Fig.~\ref{fig:Ar}b  behave predominantly $\sin \varphi$ to $-\cos\varphi$-like (blue to black area labelled '1'), the ones with $p_z<0$ behave $\cos\varphi$ to $-\sin \varphi$-like (green to red, '2'). Most of the rescattered electrons behave phase-wise similar to the direct electrons in the {\em opposite} direction ('3' red to black, and '4' blue to green). However, in the semi-circle-shaped momentum regions '5' and '6' beyond the respective $2\Up$-cutoffs  the electrons continue to follow phase-wise the direct electrons in the {\em same} direction, i.e., '2' and '1', respectively. This effect will be discussed in more detail below. Note that the overall absolute phase is not determined by the experiment so that a cyclic shift of the color code for the experimental PP spectra is permissible.

\begin{figure}
\includegraphics[width=1.0\columnwidth]{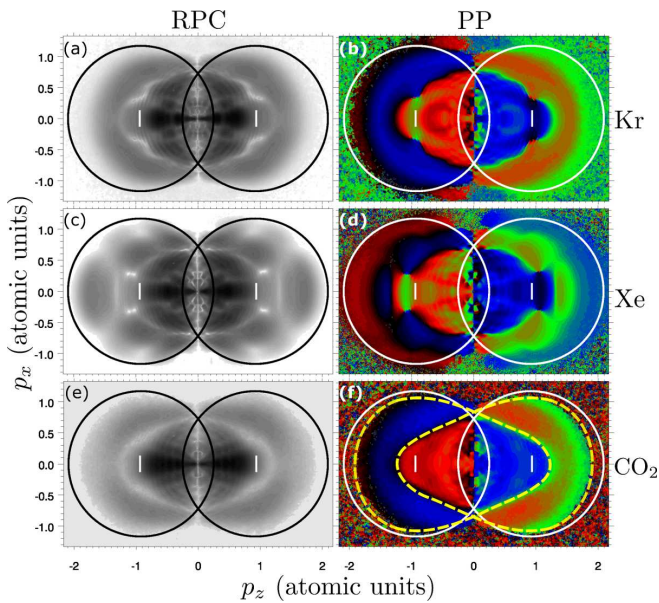} 
\caption{(color online).  Experimental RPCs (left) and PPs (right) for Kr (top), Xe (middle), and randomly aligned CO$_2$ (bottom). Laser intensities and color coding as in Fig.~\ref{fig:Ar}. The yellow dashed lines in (f) indicate the 'clubs' discussed in the text. 
 \label{fig:collection} }
\end{figure}

In Fig.~\ref{fig:collection} we show RPC and PP spectra for Kr, Xe, and randomly aligned CO$_2$. 
Figures~\ref{fig:Ar}b and \ref{fig:collection}b,d,f show that the overall structure of the PP spectra is universal:
 they resemble two overlapping clubs (indicated in Fig.~\ref{fig:collection}f), one colored in red to green (regions '2', '5', and '4' in Fig.~\ref{fig:Ar}b), the other one blue to black ('1', '6', and '3').  The blunt parts of the clubs (regions '3' and '4') represent rescattered electrons. The tip regions '5' and '6' are investigated below. 
While the overall structure is similar for all species, a target dependence is most obvious in the RPC spectra Figs.~\ref{fig:collection}a,c,e and  \ref{fig:Ar}a, and also reflected in detailed features of the PP spectra. This allows two-color spectroscopy to be employed for imaging and as a sensitive test  of theoretical models. In what follows, we aim at reproducing the results of Fig.~\ref{fig:Ar} via simple modeling using SMT.

\begin{figure}
\includegraphics[width=1.0\columnwidth]{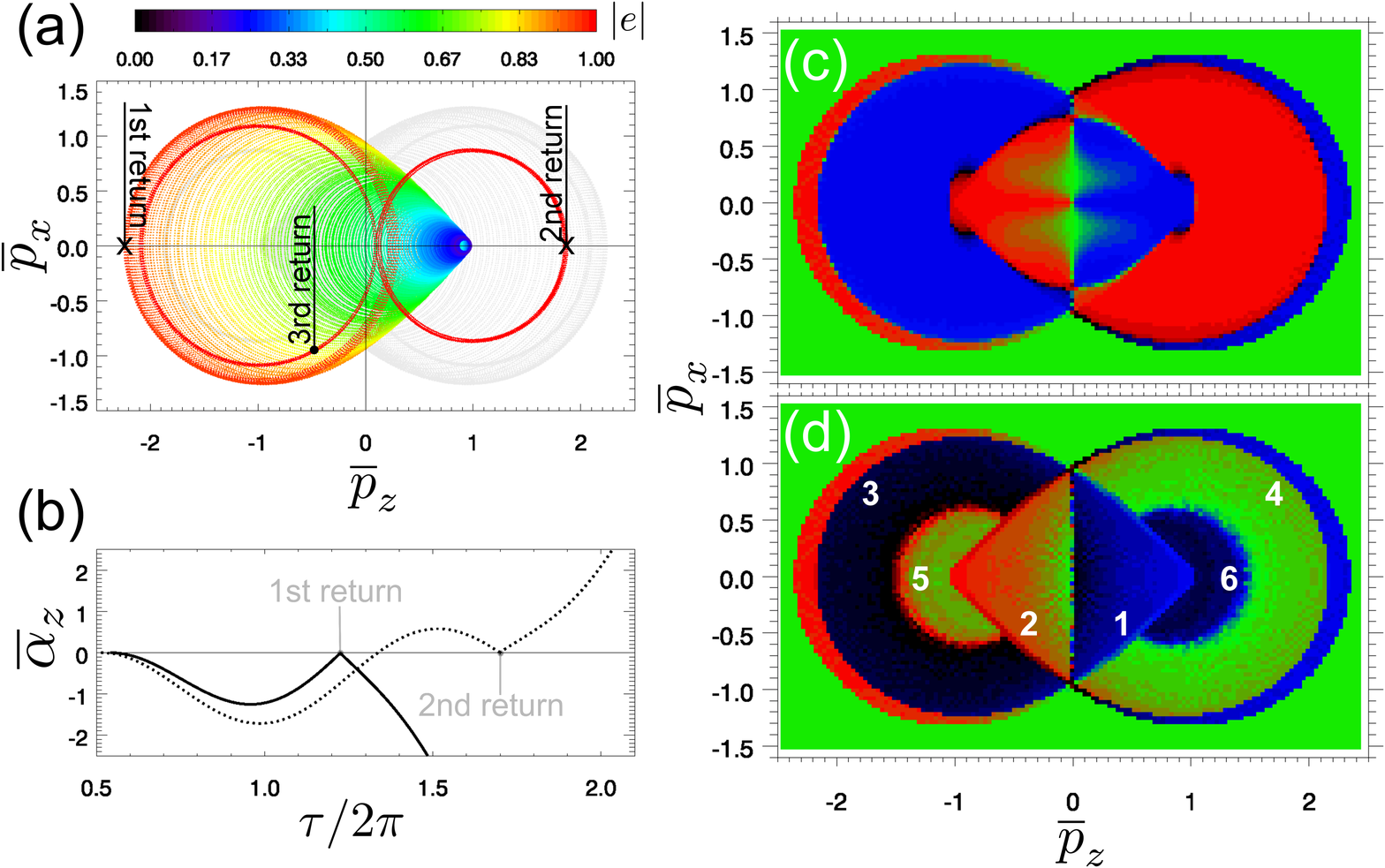} 
\caption{(color online). (a) Final SMT photoelectron momenta for $\xi=0$, absolute value of electric field at ionization time $|\vekte(\tau_1)|$ color-coded. Rings representing  local maxima in final momentum $|\overline{p}_z|$ due to rescattering during first, second and third return are labelled. For better visibility, final momenta in respective opposite directions (originating from electrons emitted half a laser cycle earlier) are plotted light gray. (b) Normalized excursion $\overline{\alpha}_z$ of electron rescattering at first (solid) and second return (dotted), respectively, leading to final momenta indicated '{\sf x}' in (a).
SMT PP spectra for $\xi=0.05$, taking into account (c) ionization and wave packet spreading ($s=3/2$) and (d) additionally a scattering cross section  \reff{eq:crosss} (with $Z=18$ and $\mu=1/2$). Same color coding in (c) and (d) as in previous figures. White numbers in (d) analogously to Fig.~\ref{fig:Ar}b.
 \label{fig:smt1} }
\end{figure}

SMT (see, e.g., \cite{ref_JPB_review,ref_quantum_orbits,ref_Scinece_Feynmann}) should be able to reveal the origins of the common, overall features observed in the PP spectra. Given a vector potential \reff{eq:vecpot} the electron momentum before rescattering reads $\vektp(\tau)/A_0 =  \overline{p}(\tau)\, \vekte_z= \vekta(\tau)-\vekta(\tau_1)$. The dimensionless time $\tau=\omega t$, momentum $\overline{p}=p/A_0$, and vector potential $\vekta=\vektA/A_0$ are introduced to highlight the universal scaling of SMT, and  $\tau_1=\omega t_1$ with $t_1$ the ionization time. Assuming isotropic scattering and considering the planar motion in, e.g., the $xz$-plane, the momentum after rescattering and once the pulse is off reads
\beq \twovec{\overline{p}_z}{\overline{p}_x} = \twovec{-a(\tau_2)}{0} + [a(\tau_2)-a(\tau_1)] \twovec{\cos\theta }{ \sin\theta } . \eeq
Here,  $\tau_2=\omega t_2>\tau_1$ with $t_2$ the rescattering time, and $\theta$ the scattering angle. The maxi\-mum momentum $\overline{p}_{\max}=\sqrt{5}$ corresponds to the well-known cut-off energy $10\Up$ which occurs if ionization happens at at $\tau_1/2\pi=0.543$ (or at $0.043$ in opposite direction) and the $\theta=\pi$-backscattering at $\tau_2/2\pi=1.22$ ($0.72$).
 The $10\Up$ rescattering rings, i.e., rings in the momentum plane centered at $-a(\tau_2)\vekte_z$ with radius $|a(\tau_2)-a(\tau_1)|$, are shown in Fig.~\ref{fig:smt1}a below and indicated in 
 all experimental and TDSE spectra. If electrons do not scatter during the first but the second return  they end up in the opposite $z$-direction and generate second-return rescattering rings. In particular, if ionization and rescattering occur at $\tau_1/2\pi=0.519$ ($0.019$)  and  $\tau_2/2\pi=1.74$ ($1.24$), respectively, a local maximum cutoff energy of  $7\Up$ is reached, also indicated in Fig.~\ref{fig:smt1}a (respective trajectory shown in Fig.~\ref{fig:smt1}b). 

 The fundamental club structure of the  PP spectra is easy to reproduce within SMT: (i) looping over $\tau_1$ and $\tau_2>\tau_1$  all electron trajectories that actually rescatter are considered, and (ii) weighted by $W=W_\mathrm{r} W_\mathrm{i}$ where $W_\mathrm{i}$ 
is an ionization and $W_\mathrm{r}$ a rescattering probability. An instantaneous tunneling rate through a triangular barrier \cite{ref:llqm} 
$ W_\mathrm{i}(\tau_1) \sim \exp\left[-{2}/{3|\vekte(\tau_1)|}\right]$ (with $\vekte(\tau)=-\partial_\tau  \vekta(\tau)$ the dimensionless electric field)
has been taken. We are not interested in absolute numbers here, as our SMT modeling should be robust and depend only on $\vekta$ and $\tau$. 
In a first, crude approximation 
$ W_\mathrm{r}(\tau_2-\tau_1)= (\tau_2-\tau_1)^{-s}$ with a spreading exponent $s>0$
has been chosen. Such a rescattering probability accounts for wave-packet spreading (see, e.g., \cite{ref_JPB_review,ref_quantum_orbits}) but neglects any momentum and angular dependence of the scattering cross section.
If several trajectories end up in the same final momentum bin only the most probable has been considered, thus neglecting interference effects. The direct electrons have been weighted by $W_\mathrm{i}^\mathrm{(dir)}(\tau_1)=W_\mathrm{i}(\tau_1) \exp(-\beta \overline{p}_x^2/|\vekte(\tau_1)|)$ with $\beta=15$ (to produce a reasonable lateral spread of the momentum distribution).  The result of such a simple modeling for $s=3/2$ is shown in Fig.~\ref{fig:smt1}c. The spreading exponent has been chosen smaller than $s=3$ for free, 3D Gaussian wave-packet spreading in order to mimic Coulomb focusing \cite{coulfocus}. The structure of the two overlapping clubs---one red, the other blue---is clearly visible. The weak  ($\xi=0.05$) $2\omega$-component leads to a small extension of the rescattering cutoff (small strip of opposite color around the blunt club ends). 

 Obvious disagreements between Fig.~\ref{fig:smt1}c and all experimental results are the oversimplified behavior consisting of essentially only two PPs (red or blue), and the too short low-momentum tip of the club ending at $|\overline{p}_z|=1$, corresponding to the $2\Up$ cut-off for direct electrons.  In all experimental and TDSE results the club tips extend beyond $2\Up$ (see regions '5' and '6' in Fig.~\ref{fig:Ar}b). 
In order to reveal the origin of this mismatch we first note that
in the SMT modeling of Fig.~\ref{fig:smt1}c the trajectories that rescatter at their first return  dominate because of the $(\tau_2-\tau_1)^{-s}$-penalty for later returns. This is the reason why there is essentially only a $\sin$-like or a $-\sin$-like behavior for all final momenta (i.e., blue or red). In a refined modeling, a rescattering probability $\sim W_\mathrm{r}(\tau_2-\tau_1) \sigma(q,\theta)$ with a differential scattering cross section (in first Born approximation)
 of the form 
\beq \sigma(q,\theta) =  \frac{(2Z)^2}{[\mu^2+ 4 q^2 \sin^2(\theta/2)]^2}, \quad q=a(\tau_2)-a(\tau_1) \label{eq:crosss}\eeq
for a screened potential $V(r)=-Z\exp(-\mu r)/r$ has been employed.  The resulting spectrum  in  Fig.~\ref{fig:smt1}d is in much better agreement with the experiments. In fact, given the simple modeling, the agreement is striking. Going from Fig.~\ref{fig:smt1}c to d, the PP in regions 3 and 4 changes from $\sin$ (blue) to $-\cos$-like (black) for $\overline{p}_z<0$, and from $-\sin$ (red) to $\cos$-like (green) for $\overline{p}_z>0$. The reason for this change lies in the introduction of an additional functional dependence on the relative phase via the cross section $\sigma[q(\varphi),\theta]$.

As in the SMT modeling all spectral features can be understood in terms of electron trajectories,  we were able to identify the origin of the ring-like extensions '5' and '6' beyond  $2\Up$. Figure~\ref{fig:smt1}a suggests that these features are due to electrons that rescatter during the second return because the corresponding rescattering rings are centered around the $2\Up$ cutoffs. A cross section of the form  \reff{eq:crosss} favors forward scattering (i.e., small $\theta$), and the more so the larger the instantaneous scattering  momentum $q=|a(\tau_2)-a(\tau_1)|$ is. As a consequence, the trajectories which scatter with lower-momentum later during the second return (incident from the opposite direction) may have a higher probability for large $\theta$ than the trajectories that rescatter during the first return, despite the wave-packet-spreading penalty for late returns.
 It is thus the competition between the probability factors governing wave packet spreading and the momentum-dependent large-angle scattering which is responsible for the club tips beyond  $2\Up$. 

\begin{figure}
\includegraphics[width=0.9\columnwidth]{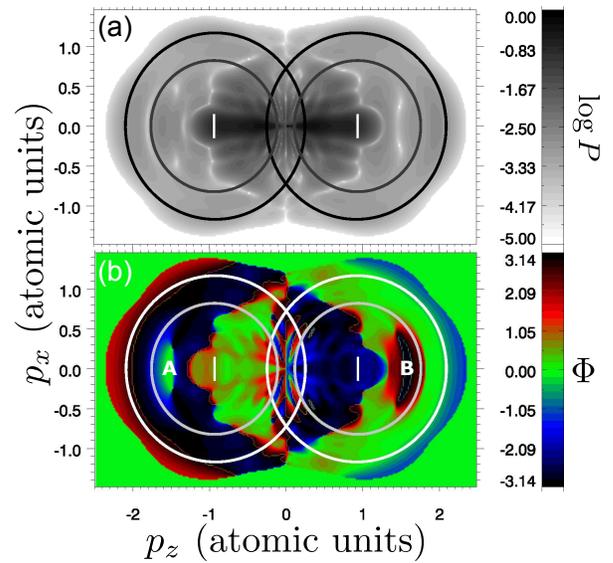} 
\caption{(color online). RPC (a) and PP (b) spectra for Ar from TDSE ($10^{14}$\,\Wcmcm, $\xi=0.05$). Durations of $\sin^2$-shaped pulses for 800 and 400\,nm were 60 and 50\,fs, respectively. Additionally to  $2\Up$ cut-offs and $10\Up$-rings, second-return cut-off rings are included, which form boundaries for features labelled 'A' and 'B'. Slight left-right asymmetries visible in (a) are remnants of CEP-dependence. 
 \label{fig:TDSE} }
\end{figure}

Figure~\ref{fig:TDSE} shows RPC and PP spectra for Ar from a TDSE simulation \cite{qprop} in single-active electron approximation. Abel-projected photoelectron momentum spectra were calculated for different $\varphi$ and subsequently treated like the experimental spectra. The correct ionization potential of Ar was imposed by choosing the 3p$_0$-orbital in the effective potential $V(r)=-[1 + 17 \exp(-2.11375 r)]/r$ as the initial state. Comparison with Fig.~\ref{fig:Ar} yields satisfactory agreement with respect to the overall club structure in the PP and the qualitative features in the RPC spectra. 
We note that the dynamic range of the
experimental detection is two orders of magnitude lower than the one shown for the TDSE. This might be the reason why the marked features close to the second-return cut-off in the TDSE PP spectrum (labelled 'A' and 'B') are absent in the experiment. The remaining differences could be due to the idealized pulses and the neglect of focal averaging in the simulation, slightly different laser parameters in experiment and simulation, or the possibly inadequate (but computationally unavoidable) assumption of a single active electron in the TDSE calculations. In any case,  such differences show the sensitivity of PP spectroscopy and its power to discriminate between various effects.

In summary, we have introduced systematic two-color spectroscopy based on RPC and PP spectra.  The presence of the $2\omega$-field component tags each emission time according to how ionization probability changes as function of relative phase. This change is subsequently mapped through the actual electron dynamics to the final photoelectron momentum. Since several trajectories end up with the same final momentum the trajectory with the largest change in general dominates. We have revealed and explained   the universal,  overall structure of the PP spectra for various rare gas atoms and CO$_2$. We further have shown that details  in both the  RPC and the PP are target-dependent, paving the way to employ them for imaging electronic structure or dynamics, and as tests for theoretical predictions or cross sections. We also anticipate that two-color spectroscopy will be able to discriminate among coherent, delayed, or thermal electron emission, e.g., in non-sequential ionization of multi-electron systems, because the relative-phase dependence is maintained, affected, or destroyed, respectively. Any spectral feature identified in an ``ordinary'' photoelectron spectrum and  hypothetically attributed to a certain phenomenon (such as rescattering or focusing of certain trajectories, polarization of the target, or internal dynamics) can be further scrutinized, in that way testing whether the PP and RPC signatures of this feature are in accordance with the predictions from the conjectured theory.

We thank M.\ Kling and S.\ Zherebtsov for bringing us in touch with the $\omega$-$2\omega$
method and helpful support in the early stage of our studies.
Fruitful discussions with M.\ Ivanov, computer time provided by the North-German Supercomputing Alliance (HLRN, project no.~mvp00004), and support through the SFB 652 of the German Science Foundation (DFG) are gratefully acknowledged.

\end{document}